\documentclass[twocol]{ametsoc}
\newcommand{\EM}[2]{
\begin{equation}
{#1}
\label{#2}
\end{equation}}
\newcommand{\E}[1]{(\ref{#1})}
\newcommand{\F}[1]{{Fig. \ref{#1}}}
\newcommand{\Fb}[1]{Figure \ref{#1}}
\newcommand{\mean}[1]{ \left\langle #1 \right\rangle }
\def\ICA{$9~\rm{km}$}
\def\ICE{$120~\rm{km}$}
\def\km{{~\rm{km}}}

\def\del{\Delta u}

\def\delul{\del_{\ell}}
\def\ulpdf{P(\delul |s)}

\def\d3{\mean{\delul^{\rm 3}}}
\def\vu{\bf u}
\def\vr{\bf s}
\def\vx{\bf x}
\def\etr{\varepsilon} 
\def\Obh{g}
\def\ce{\Obh\etr} 
\def\rms{\mean{s^2(t)}} 
\def\nms{\mean{s^n(t)}} 
\def\sigg{\rms^{1/2}} 
\title{
Inverse energy cascade in ocean macroscopic turbulence:
Kolmogorov self-similarity in surface drifter observations
 and Richardson-Obhukov constant}
    \authors{J. Dr\"ager-Dietel 
     and A. Griesel}
\affiliation{ Institut f\"ur Meereskunde, Universit\"at Hamburg, Hamburg, Germany}
\normalsize
\abstract{
We combine two point velocity and position data  from  surface drifter observations in the Benguela upwelling region off the coast of Namibia.
The compensated third order longitudinal velocity structure function $\d3/s$
shows a positive plateau  for inertial separations $s$ roughly  between \ICA~ and \ICE~ revealing 
an inverse energy cascade 
with energy transfer rate
$\etr\simeq 1.2 \pm 0.1 \cdot 10^{-7} m^3/s^2$. 
Deviations from Gaussianity of the corresponding 
probability distribution $\ulpdf$
of two-point velocity increments $\delul$
for given pair separation $s$
show up in
the n$^{th}$ antisymetric structure functions
$S_{-}^{(n)}(r)=\int u^n(P(u)-P(-u)d u$,  which scale  in agreement
with Kolmogorov's prediction, $S_{-}^{(n)}(r)\sim r^{(n/3)}$, for $n=2,4,6$.
The combination of $\etr$ with Richardson dispersion
 $\rms=\Obh\etr t^3$, where $\rms$ is mean squared pair separation at time $ t$,
reveals a  Richardson-Obhukov constant of $\Obh\simeq 0.11\pm 0.03$.
}
\begin{document} 
\maketitle
\noindent
The inverse energy cascade in (quasi) two-dimensional turbulence is one of the most important phenomena in fluid dynamics.
As described by  \citet{kraichnan1966isotropic} the coupled constraints on energy and enstrophy conservation lead to a   flux of kinetic energy from  small injection-scales towards larger scales (inverse cascade) and
a forward (down-scale) flux of enstrophy. The two cascades are reflected in a  coupled double scaling regime in the corresponding  energy spectra,
 which have been verified 
 in the last decades  by various numerical and laboratory observations.
The applicability to real geophysical flows, which, 
due to their
high Reynolds numbers,
in principle are
ideal testbeds for asymptotic turbulence theories, 
is much less explored 
but is of enormous importance.
In the ocean,  the asymptotic theory of quasigeostrophic (QG) resembles two-dimensional
(2D) turbulence  and develops the two cascades but includes density stratification, weak vertical motion, and planetary rotation \citep{charney1971geostrophic,danilov2000quasi}.
 In practice  in the ocean gridded 
satellite altimeter data \citep[e.g.][]{scott2005direct,qiu2008length}  
 currently barely resolve the Rossby Radius of deformation, which is the scale around which energy is injected by baroclinic instability. 
 Shipboard measurements, which enable to resolve smaller scales, on the other hand, yield only along-shiptrack measurements and are limited to a few regions. 
In contrast,
paired Lagrangian floats can connect a range of scales from  several 10 m
up to 1000 km and hence are a valuable source of information for 
geophysical fluid dynamics. 
\Fb{traj} shows  trajectories from drogued buoys (surface
drifters) whose  drogues   follow the ocean currents 
in $15~\rm{m}$ depth, deep in the ocean surface mixed layer of  the Benguela upwelling region, which  during the South-Atlantic summer, when the cruise took place had a depth of $30~\rm{m}$ \citep{peng2021diurnal}.
While the trajectories indicate to structures of a hierarchy of scales (see zoom in the figure), the satellite sea surface height data, represented  in the figure
by the shaded background, can only resolve scales larger than roughly $25~ \rm{km}$.
\begin{figure*}[t]
\centering
\includegraphics[width=14.0cm,angle=0]
{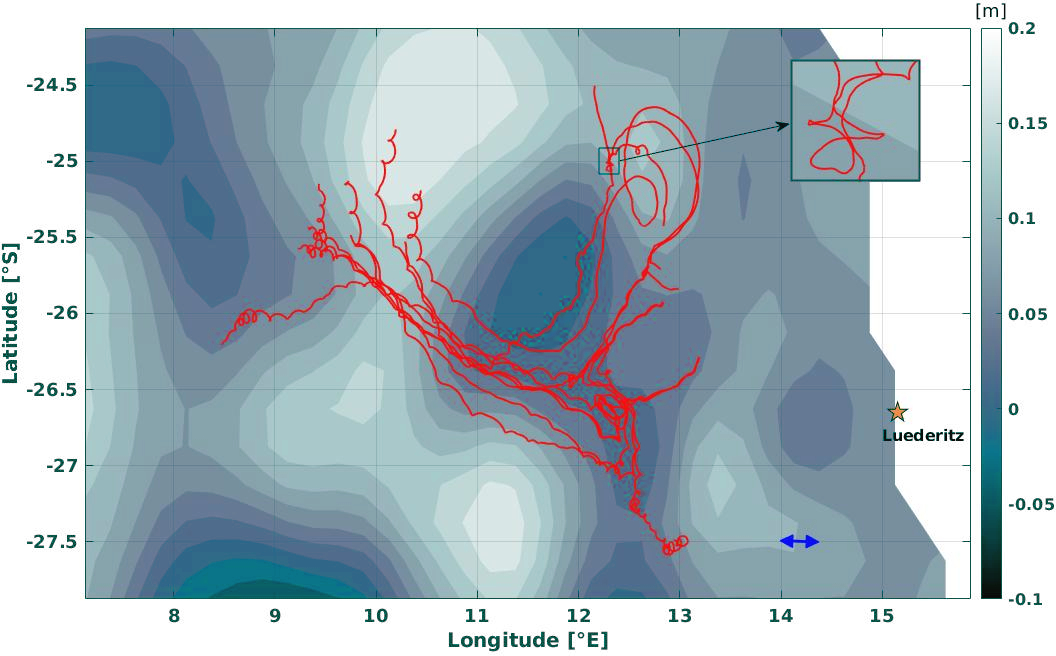}
\caption{Drifter trajectories from $5^{th}$ to $13^{th}$ day after deployment of 12
surface drifter released
in the Benguela upwelling region. 
 Inlay: Zoom into a typical  drifter trajectory  revealing a rich substructure. Background-shading is according to sea surface height.  The blue double arrow indicates the sizes of the Rossby radius $R_0\simeq 30~ \rm{km}$ of the region.
}\label{traj}
\end{figure*}

Intimately connected to the 
energy transfer between the scales is the dispersion and  mixing of tracers within the turbulent flow.
A seminal  role  in turbulent diffusion 
is played by Richardson's famous law \citep{richardson1926atmospheric},
\EM{\rms=\ce t^3,}{loc}
 which describes  the mean squared relative displacement $\rms$ of particle pairs at time $t$ in a turbulent 
flow with  $\etr$ being the energy transfer rate and $\Obh$ the Richardson-Obhukov constant. 
Numerical simulations of 3d turbulence \citep{bofetta2000inverse,ishihara2002relative} and closure models of relative dispersion \citep{sawford2001turbulent} reveal a wide range of $\Obh$-values from $0.1$ \citep{fung1998two} $- \approx 5$. 
Laboratory experiments for controlled two-dimensional turbulent flows \citep{jullien1999richardson} 
as well as for 3d turbulence \citep{ott2000experimental} 
reveal $\Obh\simeq 0.5$,
which is  by nearly one magnitude smaller than theoretically predicted by \citet{kraichnan1966isotropic}.
Besides the importance
of the Richardsons Obhukov-constant $\Obh$
 as a proof  of the underlying theory, 
it is also of fundamental quantitative relevance
as it relates the amount of pair dispersion to the  level of fluctuating energy.

In the ocean enhanced diffusion due to turbulence
 plays a crucial role in transport processes  because  turbulent diffusion is by several magnitudes larger than molecular diffusion.  Drifter experiments on Lagrangian pair
separations carried out in specific ocean regions could identify Richardson dispersion \E{loc} \citep{koszalka2009relative,lumpkin2010surface,schroeder2012targeted,sanson2017surface,drager2018relative}.
The explicit discrimination of $\Obh$ in \E{loc} 
in an  ocean region, which serves as a suitable natural testbed for quasi two-dimensional 
turbulence in the mixed layer,
is still missing despite its theoretical and practical importance.

In  this paper we fill this gap and estimate the Richardson-Obhukov constant $\Obh$
for macroscopic quasi 2D turbulence by means of a surface drifter experiment in the Benguela  upwelling region
- one of the four major Eastern boundary upwelling regions in the ocean with intense turbulent flows.
Namely, we combine the (inverse) energy-transfer rate $\etr$ derived from two point velocity data with 
Richardson's law  \E{loc} and obtain the Richardson-Obhukov constant $\Obh\simeq 0.11\pm 0.3$.
We consider data from particularly long time (120 days) drifter trajectories of
high temporal resolution (every 30 minutes),
drogued in 15 m depth in the turbulent mixed layer.
The surface drifters  (of type SVP-I-XDGS
from MetOcean Telematics  featuring Iridium telemetry)
were deployed in triplets,
placed nearly simultaneously  in  tight separations  of on average  200 m at the time of the first satellite signal. 
In \cite{drager2018relative} we focused 
on dispersion behavior by
 temporal (pair dispersion) as well as by spatial 
 measures (finite size Lyapunov exponents) both for the whole release O\{40\} and the different subsets O\{10\}.
This paper connects the energy transfer between the scales to the dispersion behavior. To this end 
we concentrate on the 
second drifter-deployment of 12 devices 
on 26 Nov 2016 which explored the upwelling region (refered to as the 'Northern release').  
Along with $\Obh$ we analyse 
whether the intrinsic asymmetry of the
distribution
of two point velocities, which is necessary to drive the inverse cascade, 
fulfills Kolmogorov's scaling behavior.

We start our analysis by estimating  the energy transfer rate $\epsilon$ from the two-point-velocity statistics. 
Following recent works 
on the scale-dependent distribution of kinetic energy \citep{balwada2016scale}
and spectral fluxes \citep{poje2017evidence}
in the Gulf of Mexico, we use
the  velocity measurements of the drifter-trajectories and  treat them as scattered point Eulerian measurements.   We  estimate
the two-point velocity increments $\delul$ for fixed separation $s$
(where the index $\ell$ denotes the longitudinal component),  defined by 
\EM{\delul=\left(\vu(\vx+\vr, t)-\vu(\vx, t)\right) \cdot\frac{\vr}{||\vr||}.}{incr}
Here, the $\delul$ 
are calculated at each time  $ t$ for each drifter pair $(i,j)$ from the
relative dispersion  matrix $D_{i,j}(t)=dist({\bf x}_i(t)-{\bf x}_j(t))$.
Statistics are obtained by time averages over the data conditioned on  binned separation distances.
The analysis in our experiment is based on roughly 6500 snapshots of the drifter-group
revealing  450 000 simultaneous  pair values.

Contrary to uncorrelated diffusive motion,
turbulence imposes correlations
between the drifter trajectories
leading to  a characteristic scaling with $s$ particularly of the   3'rd order structure functions 
$\mean{\del^3_{\ell}}$.
Specifically, Kolmogorov's
$4/5$th law for 3d-turbulence  \citep{KOLMOG41},
generalizes to any dimension d,
\EM{\mean{\del^3_{\ell}}=-\frac{12}{d(d+2)} \etr s.}{DD3gen}
While for 3d-turbulence
$\mean{\del^3_{\ell}}$ is
negative,
for 2d-turbulence $\mean{\del^3_{\ell}}$
is positive, i.e.
the  energy transfer rate $\etr$ formally has a negative sign.
The compensated third order structure function, $\d3/s$, from our drifter observations clearly shows a positive plateau 
 regime roughly  between \ICA~ and \ICE, which 
indicates an inverse energy cascade
(see \F{third} with two different refinements). 
From the plateau we derive the energy transfer rate
\EM{\etr\simeq 1.2 \pm 0.1 \cdot 10^{-7} m^3/s^2.}{etr} 

\begin{figure}[t]
\centering
\includegraphics[width=7.5cm,angle=0]
{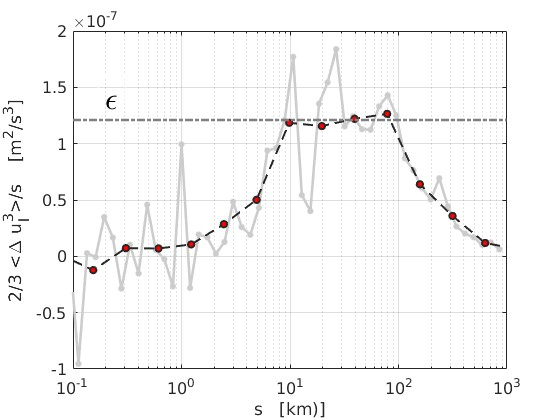}
\caption{Third order longitudinal structure function rescaled according to \E{DD3gen} for d=2 for coarse (black dashed line) and fine (light gray) bin resolutions, where the dots mark the bin centers.
The dashed gray line marks  
$\etr= 1.21   \times 10^{-7}  m^2/s^3 $.
}\label{third}
\end{figure}
\begin{figure}[b]
\centering
\includegraphics[width=5.0cm,angle=0]
{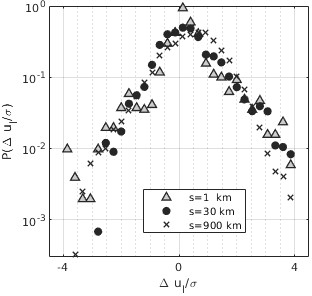}
\caption{
Rescaled probability distributions  of  longitudinal velocity increment $P(\delul/\sigma)$,
where $\sigma$ denotes the standard deviation,
for non-local (1 km), 
inertial (30 km)
and diffusive (900 km) separation scales.}\label{velpdf}
\end{figure}
\begin{figure}[t]
\centering
\includegraphics[width=8.0cm,angle=0]
{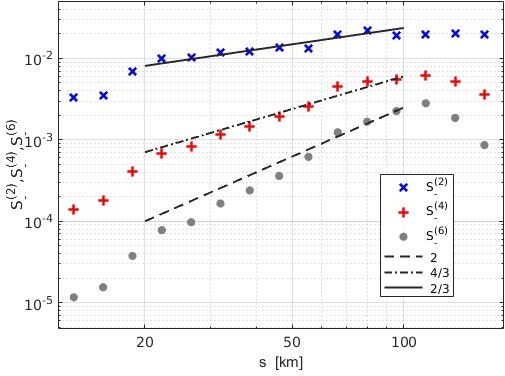}
\caption{Antisymmetric structure functions
$S_{-}^{(n)}(r)=\int u^n(P(u)-P(-u)d u$ of order 
$n=2,4,6$. The straight lines represent the theoretical slopes according to
Kolmogorov's prediction $S_{-}^{(n)}(r)\sim r^{(n/3)}$. 
}\label{antipdf}
\end{figure}
\begin{figure}[b]
\centering
\includegraphics[width=8.0cm,angle=0]
{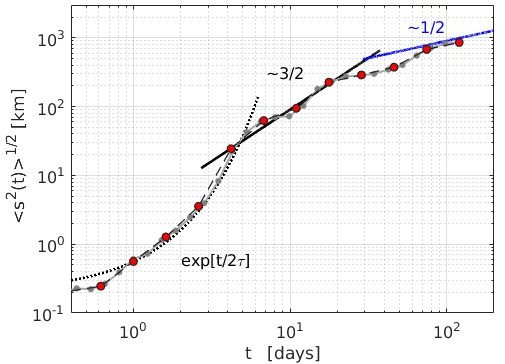}
\caption{Root mean square pair separation $\sigg$  for drifter pairs
plotted against time $ t$
 shows the characteristic distinct dispersion regimes
   (non-local (dotted), Richardson \E{loc} (full black line) and diffusive (blue line)) of an ocean surface mixed layer. The corresponding Richardson pdf was  
   analysed in \cite{drager2018relative} by the scaling behavior 
   of additional moments $\nms$.}\label{disp}
\end{figure}
The correlations in the velocity field, which lead to \E{DD3gen}, 
are accompanied by
asymmetries of the corresponding 
probability distribution $\ulpdf$ 
of two-point velocity increments $\delul$
for given pair separation $s$, which are necessary to drive the inverse cascade.
Because (quasi) 2d-turbulence, in contrast to 3d-turbulence,
is expected to be non-intermittent 
 visually only subtle deviations from Gaussianity are expected for $\ulpdf$ \citep{bofetta2000inverse}.
In \F{velpdf} we  plotted
the distribution $p(\delul/\sigma)$ of rescaled two-point velocity increments  $\delul/\sigma$ for 
3 separation-bins $s$ 
corresponding to the 
three dynamical regimes, the pre-inertial (non-local), the inertial and the diffusive regime.
Except for pre-inertial scales ($s=1\km$, triangles), where 
$P(\delul/\sigma)$ shows  clear deviations from gaussianity in the core region,
negative or positive skewness 
can hardly be detected from visual impressions.
To clearly access the tails of the distributions
and possible asymmetries 
on more quantitative grounds we follow \cite{bofetta2000inverse} and 
analyze the n$^{th}$ antisymmetric structure functions
$S_{-}^{(n)}(r)=\int u^n(P(u)-P(-u)d u$.
 According to \F{antipdf}, the $S_{-}^{(n)}$
of order 
$n=2,4,6$ show a scaling behavior in agreement with Kolmogorov predictions 
$S_{-}^{(n)}(r)\sim r^{(n/3)}$ 
represented in the figure by the straight lines.
Because  with increasing order $n$ the moments $S_{-}^{(n)}$ focus on 
outer tails of $\ulpdf$,
this indicates that the non-Gaussian  antisymmetric part, although visually small, has imprinted all the relevant scaling information on the inverse cascade.

Finally, we combine our results for the
$3^{rd}$ order relative (longitudinal)  velocity structure function  $\mean{\del^3_{\ell}}$
 (\F{third}) 
 with the  mean squared pair separation $\rms$).
 The data support  Richardson's  law \E{loc} with
  $\ce\simeq 1.3 \pm 0.2 \cdot10^{-8} m^3/s^2$ 
    (with error from compensated plot, not shown)
 for nearly a decade from day 4 on (\F{disp}). 
 Combination of  $\ce$  with the energy transfer rate $\etr$, see \E{etr}, 
reveals for the dimensionless Obhukov constant
$\Obh$ the following estimate:
\EM{\Obh\simeq 0.11 \pm 0.03.}{RiOb}
To conclude, in this paper, we investigate the inverse cascade by  
a Semi-Lagrangian analysis of real ocean
two point velocity and position data, namely from  surface drifter observations 
in the Benguela upwelling region.
The compensated $3^{rd}$ order relative (longitudinal)  velocity structure function $\mean{\del^3_{\ell}}/s$  
 reveals a positive plateau for inertial scales roughly between  \ICA~ and \ICE~ 
 indicating an inverse energy cascade with 
 an energy transfer rate 
 $\etr\simeq 1.2 \pm 0.1 \cdot 10^{-7} m^3/s^2$,
 which is very close to the one measured with $(\sim 1~m)$ deep drogued drifters 
 in the northern Gulf of Mexico from
 the GLAD (Grand LAgrangian Deployment) observational program  \citep{balwada2022direct}. 
Deviations
from Gaussianity of the corresponding 
probability distribution $\ulpdf$
for inertial $s$
show up in the n$^{th}$ antisymmetric structure functions
$S_{-}^{(n)}(r)=\int u^n(P(u)-P(-u)d u$ which
show a scaling behavior in agreement with Kolmogorov's prediction,
 $S_{-}^{(n)}(r)\sim r^{(n/3)}$, for $n=2,4,6$ (similarly to what was found in \cite{bofetta2000inverse}). 
This indicates that the antisymmetric part of $P(\delul)$, although visually small, has imprinted all the relevant scaling information on the inverse cascade.
For smaller  scales, 100\,m-10\,km, 
(\citet{poje2017evidence,essink2019can}),  also from GLAD,
found evidence of a forward energy cascade, namely negative $\mean{\del^3_{\ell}}/s$
with transfer rate of $\etr\simeq 2.0 \pm 0.1 \cdot 10^{-7} m^3/s^2$ 
and left skewed $P(\Delta u_{\ell} |s)$. 
The combination with two point position data from the drifters revealed  a Richardson constant of $\Obh \sim 10^{-2}$ for the direct cascade.
The apparent reduction in pair dispersion for a given level of fluctuating energy was suggested to  stem from a high degree of horizontal
 convergence  in the surface velocity field.
In contrast, for the inverse cascade considered here, the combination of $\etr$  gained from
the $3^{rd}$ order relative velocity structure function  $\mean{\del^3_{\ell}}$  with the mean squared  pair separation $\rms=\Obh\etr t^3$
  reveals  an Obhukov constant of $\Obh\simeq 0.11 \pm 0.03$.
   Our results thereby provide the important link between laboratory and field observation
   as well as numerical simulation.

We thank Dhruv Balwada for valuable discussions and Program-code from \citep{balwada2016scale} in the initial phase of the project.
This paper is a contribution to the Collaborative Research Centre TRR 181 ”Energy Transfers in Atmosphere and Ocean” funded by the Deutsche Forschungsgemeinschaft (DFG, German Research Foundation) - project number 274762653.

\bibliographystyle{ametsoc2014}

\bibliography{referfield3}

\end{document}